\documentclass[aps,pre,preprint,12pt,superscriptaddress,showkeys]{revtex4}
\usepackage{graphicx}
\usepackage{amsmath,amssymb}
\begin{document}
\title{Historic Trends in U.~S. Drought Forcing in a Warming Climate}
\author{T.~Muschinski}
\affiliation{Department of Physics\\Washington University, St. Louis, Mo.
63130}
\author{J.~I.~Katz}
\affiliation{Department of Physics and McDonnell Center for the Space
Sciences\\Washington University, St. Louis, Mo. 63130}
\email{katz@wuphys.wustl.edu}
\begin{abstract}
The mean North American and world climates have warmed since the beginning
of climatologically significant anthropogenic emission of greenhouse gases
in the 19th Century.  It has been suggested that warming may increase the
frequency or severity of droughts.  We define and study the statistics of an
aridity index that describes the precipitation forcing function of drought,
considering drought to be a season with low enough precipitation to be
significant for agriculture.  Our aridity index is a reciprocal function of
the seasonal precipitation, which is more significant for agriculture than
mean precipitation.  Using NOAA data from sites in 13 diverse climate
regimes in the 48 contiguous United States with time series running over the
period 1940--1999 but including two data series from 1900 or 1910, and
computing their decadal averages, we search for linear trends in their
aridity indices.  We find no linear trends significant at the $2\sigma$
level.  At five sites $3\sigma$ upper bounds on any systematic trends are in
the range 1.0--2.8\%/decade, while at two sites $3\sigma$ lower bounds are
-0.5\%/decade and -2.2\%/decade; at other sites the bounds are less
restrictive. 
\end{abstract}
\maketitle
\section{Introduction}
\label{sec:Introduction}
The mean world climate has warmed \cite{JM03,MW05,TJ07,HRSL10,BEST} since
the the beginning of climatologically significant anthropogenic emission of
greenhouse gases in the 19th Century. It has been suggested 
\cite{IPCC1,IPCC2,CR12} that warming is accompanied by increases in the
frequency of ``extreme weather events'', a broad category that includes
severe storms and drought.  Conclusions drawn from single extreme events
are controversial\cite{HSM13}, but long term averages carry more statistical
power.

There are many ways of defining ``extreme events'', and it is necessary
to find objective quantitative measures.  It is difficult to predict future
changes in the hydrological cycle from climate models \cite{AI02}, but
guidance may be found in the historic record of climate as it has warmed
over the last century.  In this paper we are concerned with the periods of
low precipitation that are the forcing function of drought.

Drought has been a concern of humanity since the prehistoric development of
agriculture.  It is a complex phenomenon that may be defined in many ways.
For example, the widely used Palmer Drought Index (PDI) and the Palmer
Drought Severity Index (PDSI) \cite{P65,K86,D11a,D11b} involve a complex
interplay among precipitation and modeled evapotranspiration, soil moisture,
runoff and recharge (but do not include the effects of humidity, vegetation,
cloud cover, precipitation rate, wind and soil permeability).  The PDI and
PDSI are useful to agriculturalists and water resource engineers because
they measure the deviation of local conditions from their long-term means
that are the basis of planting and planning.  These indices filter the
precipitation forcing through a complex and model-dependent transfer
function.  To consider the possible effects of climate change on drought we
separate the forcing by precipitation from other processes, even while
acknowledging that these other processes (such as temperature change, which
affects the evapotranspiration rate) contribute to the response of the
hydrological system.

A number of other drought indices exist \cite{Q09,D11b}, but are also
imperfect tools for studying the possible effects of climate change on
precipitation.  For example, the Standardized Precipitation Index (SPI)
\cite{MK93,MK95,G98,G99} compares the precipitation at a site over some
period (chosen in the range from one month to several years) to the
statistical distribution of recorded precipitation at that site in periods
of that length.  This identifies anomalous (unusually dry or wet) periods
that are then assigned a quantitative index value based on the fraction of
such periods in the record that were dryer or wetter.  The SPI describes
how unusual is the value of precipitation at that site (without making the
unproven assumption of a Gaussian distribution), rather than quantifying
the magnitude of its deviation from the mean or its implications.

Many previous studies of long term precipitation trends have been concerned
with total annual precipitation.  This is a measure of climate change that
is not directly applicable to drought; a dearth of precipitation over a few
months of a growing season is drought to the farmer, even if the annual
total is high.

Studies of drought trends using the Palmer Drought Severity Index (PDSI;
\cite{KH90,DTK98,DTQ04,D11a}) are inconsistent, with the earliest work
finding no evidence of a trend but some recent work \cite{D11b,D13}
indicating a drying trend during a period that mostly overlaps with that
considered here.  Because the PDSI includes soil drying as a result of
increased evapotranspiration as the climate warms, it conflates effects of
precipitation and temperature changes \cite{SWR12} and is not a direct
measure of the precipitation forcing function.  Warming increases
evapotranspiration and biases the PDSI towards drought.  Our purpose is to
determine or constrain directly any historic trend in the precipitation
forcing function.

Longer term studies \cite{CWEMS04,D11b} suggest a correlation between
proxy drought measurements in North America and the (northern European)
Medieval Warm Period.  Their applicability to the modern period of warming
by greenhouse gases is uncertain.

Studies using the SPI may be more directly comparable to ours, but are few.
For example, \cite{GR03} modeled drought in the northeastern United States
and \cite{LBS10} analysed SPI data for Hungary, but neither of these
publications present detail sufficient for comparison.

The purpose of this work is to determine if extended periods of low
precipitation that are the forcing function of droughts have become more (or
less) frequent or severe in the 48 contiguous United States as the climate
has warmed in the last century.  We wish to separate changes in
precipitation from those of temperature with which they are conflated in
drought indices.  Dearths of precipitation are basic and elemental
parameters of climate change.  We ask if the frequency and severity of
periods of low precipitation have changed.  This question has comparatively
simple and unambiguous statistical measures, free of the complications,
inherent in drought indices, of including parameters, such as humidity,
vegetation, cloud cover, precipitation rates and wind, for which data may be
absent or limited, but that affect evapotranspiration and runoff.

In order to avoid the subtleties and complications of modeling \cite{D13}
our approach is entirely empirical, and we forgo any attempt to interpret
these historical results as tests of the validity of climate models or of
their predictions of drought.  Nor do we attempt to separate secular or very
long term (on time scales of 50 years or more) trends from natural
variability on shorter time scales.  Because of the ``red'' spectrum of
natural climatic variation \cite{MW69,P98} and its complex dependence on
space, time and the variable considered \cite{PSH09}, we do not attempt the
difficult task \cite{HEP10} of separating long term natural variations from
gradual anthropogenic climate change.

Here we define an empirical aridity index that measures any seasonal dearth
of precipitation, compute its decadal averages, and determine or bound any
long term trends.  We consider 13 sites in distinct climatic zones within
the 48 contiguous United States, with data records from 1940--1999 for most
sites, but with two extending back to the first decade of the 20th Century.
Because drought conditions are generally regional, these comparatively few
sites sample the climate of a large area, including most North American
climate zones.  From these data we are able to bound the historic rate of
change of the precipitation forcing function of drought.
\section{Methods}
\label{sec:Methods}
We use a NOAA hourly precipitation database \cite{NOAA} to construct
precipitation totals $P_{i,j}$ for the three-month periods, approximately
corresponding to the seasons, January--March, April--June, July--September
and October--December, where $i$ denotes the site and $j$ the calendar
quarter and year.  We define the annual mean seasonal aridity index:
\begin{equation}
A_{i,Y} \equiv \frac{1}{4} \sum\limits_{j \in Y} \frac{1}{(P_{i,j} + C)^2},
\end{equation}
where $Y$ denotes the year.  This index is strongly influenced by the
severity of dry periods, and much less by variations in precipitation in wet
periods.  This metric differs from the frequently used \cite{D11b,D13} mean
annual precipitation; the agriculturalist is chiefly concerned with dry
seasons that are hardly mitigated by intervening wet periods.

$A_{i,Y}$ is regularized by the addition of the constant $C$ to the
denominator, avoiding a singularity if there is no precipitation at all.  We
take $C = 6^{\prime\prime} = 15.24$\,cm so that agriculturally
insignificant precipitation has little effect on a quarter's contribution to
$A_{i,Y}$.  In order to avoid bias resulting from the omission of a season
(that might be seasonally dry or wet) when only incomplete data are
available, $A_{i,Y}$ is not computed if precipitation data are not available
for the entire year, and that year is excluded from the analysis.

We define the decadally averaged mean seasonal aridity index:
\begin{equation}
A_{i,D} \equiv \frac{1}{N_{i,D}} \sum\limits_{Y \in D} A_{i,Y},
\end{equation} 
where $N_{i,D}$ is the number of years with complete data in the decade $D$.
If fewer than five years are present $A_{i,D}$ is not computed.  The 
uncertainty of $A_{i,D}$ is estimated:
\begin{equation}
\sigma_{i,D} \equiv \frac{1}{N_{i,D}} \sqrt{\sum\limits_{Y \in D} (A_{i,Y} -
A_{i,D})^2}.
\end{equation}
\section{Results}
\label{sec:Results}
\begin{figure}[h!]
\begin{center}
\includegraphics[width=4.5in]{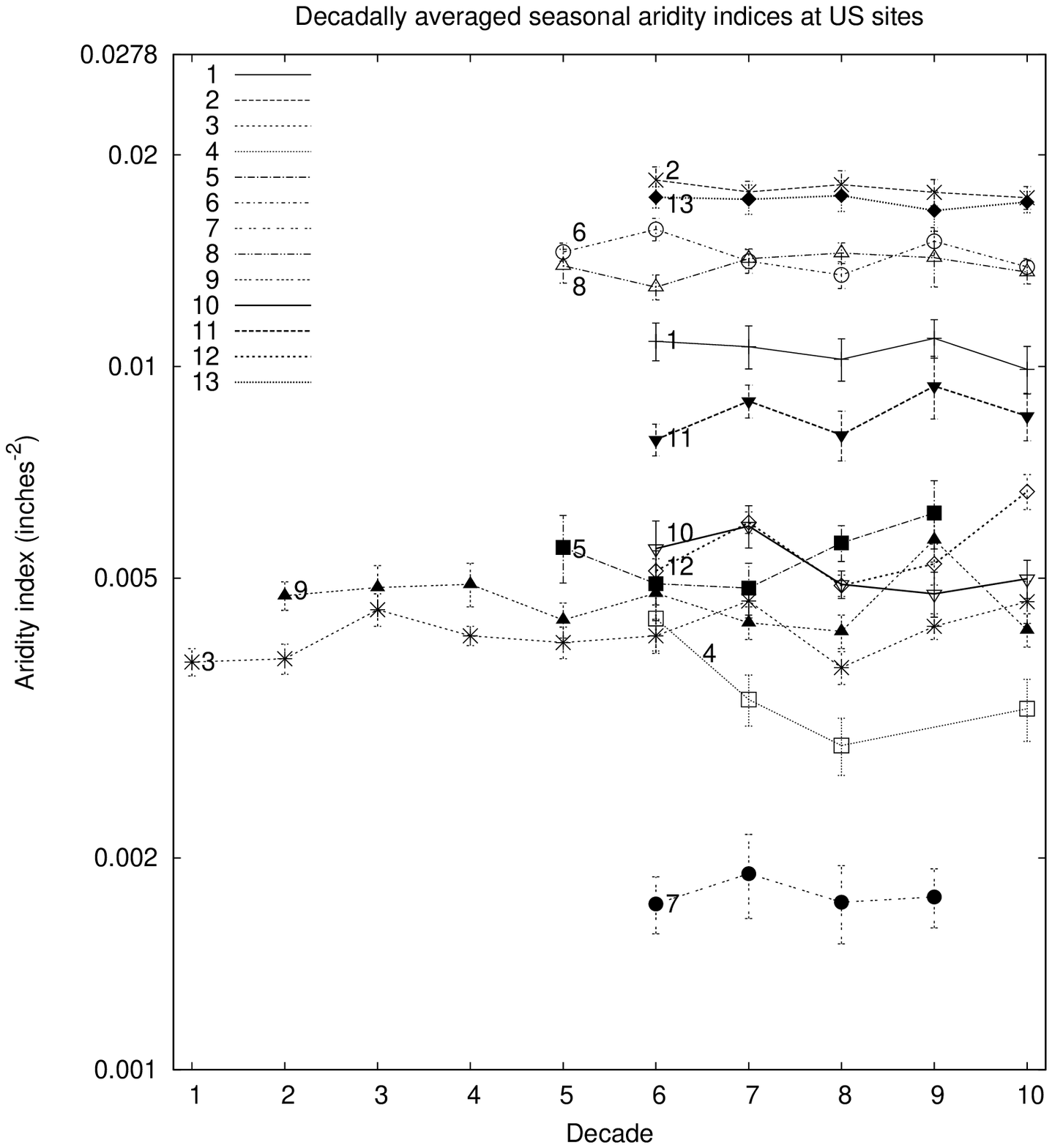}
\end{center}
\caption{Aridity indices $A_{i,D}$ at 13 sites in the 48 contiguous
United States.  Error bars are $\pm 1\sigma$.  Decade 1 is 1900--09, {\it
etc.\/}  The maximum possible value of $A_{i,D}$ is 0.0278/in$^2$.}
\label{droughtfig}
\end{figure}

The $A_{i,D}$, with error bars $\pm \sigma_{i,D}$, are plotted in
Fig.~\ref{droughtfig}.  Specifications of the sites, $\chi^2$ of the
no-trend (null) hypothesis and parameters of the best fit linear trends are
shown in Table \ref{droughttable}.
\begin{table}[h!]
\begin{center}
\begin{tabular}{|c|c|l|c|c|r|c|}
\hline
Key & NOAA Site &\qquad\qquad Location\ & Lat.\,(N) & Long.\,(W) & $\chi^2$\,(d.o.f.) & Slope \\
\hline
 1 & 140620 & Bazine 13 mi SSW, KS & $38^\circ\,16^\prime$ & $\phantom{0}99^\circ\,45^\prime$ & 1.49 (4)\ & $-1.4 \pm 2.2$ \\
 2 & 020080 & Ajo, AZ & $32^\circ\,22^\prime$ & $112^\circ\,52^\prime$ & 1.17 (4)\ & $-1.1 \pm 1.3$ \\
 3 & 366889 & Philadelphia Airport, PA & $39^\circ\,52^\prime$ & $\phantom{0}75^\circ\,14^\prime$ & 19.80 (9)\ & $+1.0 \pm 0.5$ \\
 4 & 010008 & Abbeville, AL & $31^\circ\,25^\prime$ & $\phantom{0}85^\circ\,17^\prime$ & 8.39 (3)\ & $-5.5 \pm 3.2$ \\
 5 & 081271 & Canal Point Gate 5, FL & $26^\circ\,52^\prime$ & $\phantom{0}80^\circ\,38^\prime$ & 4.35 (4)\ & $+3.9 \pm 2.7$ \\
 6 & 241088 & Bredette, MT & $48^\circ\,33^\prime$ & $105^\circ\,16^\prime$ & 10.40 (5)\ & $-1.2 \pm 0.8$ \\
 7 & 450013 & Aberdeen 20 mi NNE, WA & $47^\circ\,16^\prime$ & $123^\circ\,42^\prime$ & 0.37 (4)\ & $+0.3 \pm 4.9$ \\
 8 & 047633 & Sacramento, CA & $38^\circ\,25^\prime$ & $121^\circ\,30^\prime$  & 5.50 (5)\ & $+0.5 \pm 1.0$ \\
 9 & 310301 & Asheville, NC & $35^\circ\,35^\prime$ & $\phantom{0}82^\circ\,33^\prime$ & 17.42 (8)\ & $-1.1 \pm 0.7$ \\
10 & 111577 & Chicago Midway Airport, IL & $41^\circ\,44^\prime$ & $\phantom{0}87^\circ\,47^\prime$ & 6.61 (4)\ & $-3.8 \pm 2.0$ \\
11 & 217294 & St. Cloud, MN & $45^\circ\,33^\prime$ & $\phantom{0}94^\circ\,03^\prime$ & 4.34 (4)\ & $+2.0 \pm 2.1$ \\
12 & 431081 & Burlington, VT & $44^\circ\,28^\prime$ & $\phantom{0}73^\circ\,09^\prime$  & 30.03 (4)\ & $+2.0 \pm 1.4$ \\
13 & 045114 & Los Angeles Int. Airport, CA & $33^\circ\,56^\prime$ & $118^\circ\,24^\prime$ & 0.46 (4)\ & $-0.5 \pm 1.4$ \\
\hline
\end{tabular}
\end{center}
\caption{Sites and results.  The penultimate column gives $\chi^2$ and
(number of degrees of freedom) for the hypothesis of a constant aridity
index $A_{i,D}$ equal to its uncertainty-weighted mean.  The last column
gives the best fit linear trend of $A_{i,D}$ and its $\pm 1\sigma$
uncertainty in \%/decade.}
\label{droughttable}
\end{table}

As expected, the aridity index is largest at the desert site 2 (its maximum
possible value is $(6^{\prime\prime})^{-2} = 0.0278^{\,\prime\prime^{-2}}$),
and almost as large at Californian
sites 8 and 13 where summer precipitation is rare.  Next largest are high
Plains sites 6 and 1, and then more easterly sites with more summer
precipitation.  Finally, the aridity index has its lowest values at site 7
on the Olympic peninsula, with year-round frequent light precipitation.

We find no linear trends significant at the $2\sigma$ level.  At five sites
the $3\sigma$ upper bounds on any such trends are in the range 
1.0--2.8\%/decade, while at two sites the $3\sigma$ lower bounds are
-0.5\%/decade and -2.2\%/decade; at other sites the bounds are less
restrictive.   This result is consistent with the prediction \cite{M12}
that a $1.4^{\circ\,}$C warming, nearly twice the warming since the
beginning of the industrial era (and an even greater multiple of the warming
over the span of our data), will be required for precipitation changes
to become statistically significant ({\it N.B.:\/}  These authors are
concerned with wet season precipitation, while our aridity index measures the
driest seasons of the year, so these results, though consistent, are not
strictly comparable.). 

Our uncertainty estimates assume a normal distribution.  While it is known
(and obvious) that short-term precipitation statistics are far from
Gaussian, these estimates only assume that the distribution of annual values
within a decade of the aridity index is Gaussian, a plausible (though
unproven) assumption because many weather systems contribute to a seasonal
or annual precipitation total.  Because of the existence of long-time
correlations (``red'' noise) in geophysical data \cite{MW69,P98}, it is likely
that the tails of the distributions are greater (``fatter'') than those of
normal distributions.  If we had found apparently significant trends, this
would reduce their statistical significance; here it weakens (to a degree
that cannot be calculated because we do not know the true distributions)
the bounds that can be placed on trends.  It does not weaken our null result
that no significant trends can be found in the data.

Despite the absence of significant linear trends, the constant hypothesis is
rejected by the $\chi^2$ test at the $P < 0.02$ level at site 3, at the
$P < 0.03$ level at site 9, and at the $P < 0.001$ level at site 12.  This
reflects the well-known fact that there are long-period (decadal and longer)
variations in the climate system \cite{MW69,H76,M97,P98,MH02}, so that
decadal means may differ significantly from longer-term means even in the
absence of a linear trend.

\appendix
\section{Data}
\label{ref:MM}
We use precipitation data from \cite{NOAA}, summing the rainfall in each
quarter.  When hourly data are missing we use the cumulative data.  We
searched the data for anomalous values, such as negative values or values
in excess of $5^{\prime\prime}$ in one hour.  Such large values might
indicate suspect data because the all-time record hourly rainfall in the 48
contiguous United States is $8^{\prime\prime}$, and most state all-time
hourly records are in the range 5--7$^{\prime\prime}$ \cite{MR02}.  We found
only one such instance in our database, an hourly value of more than
10$^{\prime\prime}$ that was inconsistent with the daily total of less than
2$^{\prime\prime}$ and that we discarded, using the daily value.

\begin{figure}[h!]
\begin{center}
\includegraphics[width=4.5in]{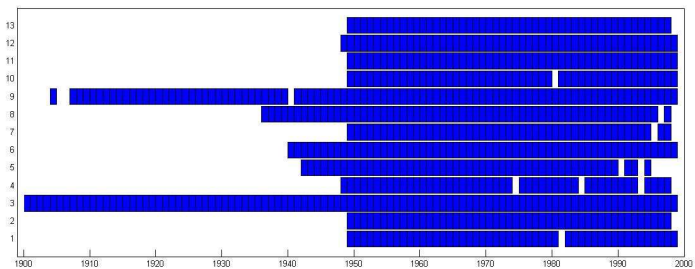}
\end{center}
\caption{Data coverage, showing years with complete data at our 13 sites.
Decadal averages are only computed if there are five or more years of data
in the decade.}
\label{datacoverage}
\end{figure}

Data were included from years for which the data sets were complete.  These
years are indicated in Fig.~\ref{datacoverage}.  Decadal means and standard
deviations (obtained from the standard deviations of the aridity indices for
that site within the decade considered) were computed only if data were
available for at least five years in the decade; other decades were omitted.

In order to check for data homogeneity, we calculated the run statistics
\cite{B68} of deviations from their means of the annual aridity index for
each site and from the linear fits to these annual data; these two
statistics gave essentially identical results.  We found no statistically
significant deviations from homogeneity at twelve of our sites, but at site
12, where the constant hypothesis is rejected at the $P < 0.001$ level by
the $\chi^2$ test, inhomogeneity was significant at the 98\% confidence
level.  These results are not independent; they reflect roughly decadal
variations that appear both as significant deviations of decadal means from
a constant value and as long runs of positive or negative annual deviations
from the mean (and hence fewer distinct runs than for homogeneous data),
despite the absence of a significant linear trend.  This result may be
interpreted as the consequence of natural variability, as was found for
Scottish rainfall \cite{H03}.  However, finding one such result in 13
independent data series is only significant at the 75\% level.

%
%

\begin{acknowledgments}
We thank Novim for support.
\end{acknowledgments}


\bibliography{droughtarxiv}   

%
%

\end{document}